\documentclass[prb,twocolumn,showpacs,amsmath,amssymb,preprintnumbers]{revtex4}
\usepackage{dcolumn}
\usepackage{bm}
\usepackage{graphicx}
\usepackage{color}

\begin{document}

\title{Insulating phase in Sr$_2$IrO$_4$: An investigation using critical analysis and magnetocaloric effect}

\author{Imtiaz Noor Bhatti}\affiliation{School of Physical Sciences, Jawaharlal Nehru University, New Delhi - 110067, India.}
\author{A. K. Pramanik}\email{akpramanik@mail.jnu.ac.in}\affiliation{School of Physical Sciences, Jawaharlal Nehru University, New Delhi - 110067, India.}

\begin{abstract}
The nature of insulating phase in 5$d$ based Sr$_2$IrO$_4$ is quite debated as the theoretical as well as experimental investigations have put forward evidences in favor of both magnetically driven Slater-type and interaction driven Mott-type insulator. To understand this insulating behavior, we have investigated the nature of magnetic state in Sr$_2$IrO$_4$ through studying critical exponents, low temperature thermal demagnetization and magnetocaloric effect. The estimated critical exponents do not exactly match with any universality class, however, the values obey the scaling behavior. The exponent values suggest that spin interaction in present material is close to mean-field model. The analysis of low temperature thermal demagnetization data, however, shows dual presence of localized- and itinerant-type of magnetic interaction. Moreover, field dependent change in magnetic entropy indicates magnetic interaction is close to mean-field type. While this material shows an insulating behavior across the magnetic transition, yet a distinct change in slope in resistivity is observed around $T_c$. We infer that though the insulating phase in Sr$_2$IrO$_4$ more close to be Slater-type but the simultaneous presence of both Slater- and Mott-type is the likely scenario for this material.
\end{abstract}

\maketitle
\section{Introduction}
In recent times, lot of scientific interests have been placed on Ir-based 5$d$ transition metal oxides (TMOs). These materials have enhanced spin-orbit coupling (SOC) effect and sizable crystal field effect (CFE) due to presence of heavy Ir atoms. On the other hand, the electronic correlation effect ($U$), which has reasonable effect in 3$d$ based TMOs, is much weak in these materials due to extended character of 5$d$ orbitals. The complex interplay between these competing interactions such as, SOC, $U$ and CFE is believed to give rise $J_{eff}$ = 1/2 electronic state which in turn stabilizes various exotic ground states in these Ir-based oxide materials.\cite{kim1}

The structural organization in iridium oxides plays crucial role for their magnetic and electronic states. Following general Ruddlesden-Popper series Sr$_{n+1}$Ir$_{n}$O$_{3n+1}$, the $n$ = 1 realizes layered compound Sr$_2$IrO$_4$ which is highly insulating and has canted-type antiferromagnetic (AFM) state at low temperature.\cite{crawford} The $n$ = 2 gives bi-layered compound Sr$_3$Ir$_2$O$_7$ which also has an insulating state and antiferromagnetic type spin ordering at low temperature, however, resistivity shows much moderate value compared to its $n$ = 1 counterpart.\cite{cao-bi} In contrast, the SrIrO$_3$ which is stabilized for $n$ = $\infty$, shows a metallic ground state with paramagnetic behavior.\cite{moon} Nonetheless, insulating phases in iridium oxides appear to be related with the nature of magnetic state of the materials. Recently, an intriguing coupling between structure, magnetism and electrical conductivity has been shown for Sr$_2$IrO$_4$.\cite{imtiaz} Therefore, understanding the nature of magnetism is much necessary to comprehend the nature of insulating phase in these materials.

The K$_2$NiF$_4$-based layered compound Sr$_2$IrO$_4$ has drawn special interest as recent theoretical calculations have predicted for possible superconductivity in this material due to its structural similarity with superconducting La$_2$CuO$_4$ and Sr$_2$RuO$_4$ materials.\cite{wang,yang,hiroshi,gao} On cooling, Sr$_2$IrO$_4$ exhibits paramagnetic (PM) to canted antiferromagnetic (AFM) transition around 230 K. The successive rotation of IrO$_6$ octahedra ($\sim$11.3$^o$ along c-axis) induces Dzyaloshinsky-Moriya (DM) type antisymmetric exchange interaction which in turn gives a weak ferromagnetic (FM) behavior below the magnetic ordering temperature. The insulating state in Sr$_2$IrO$_4$ is rather much debated, and has been largely explained with the description of Mott's and Slater's theory. Following the former picture, the 5$d^5$ electronic state of Ir$^{+4}$ completely fills the $J_{eff}$ = 3/2 and partially fills the $J_{eff}$ = 1/2 state. The $J_{eff}$ = 1/2 has very narrow bandwidth which in presence of even small $U$ splits up, and opens a Mott-like gap.\cite{kim1,kim2,jackeli} In this picture, insulating phase is driven by an electronic correlation effect being independent of appearance of magnetic ordering though this model favors the localized picture of spin interaction. In the picture of Slater model, on other hand, insulating phase is a consequence of magnetic transition as the long-range type antiferromagnetic ordering results in opening of band gap at the boundaries of folded Brillouin zone.\cite{gebhard} This branch of insulator basically has mean-field type magnetic interaction with itinerant character.

Both theoretical as well as experimental evidences have been put forward in support of both these insulating phases in Sr$_2$IrO$_4$. The $J_{eff}$ = 1/2 dominated Mott insulating phase has been discussed with the theoretical models\cite{jackeli} and also been verified from experiments such as, angel resolved photoemission spectroscopy (ARPES)\cite{kim1} and resonant x-ray scattering (RXS)\cite{jungho} experiments. In parallel developments, a recent study by x-ray absorption spectroscopy have shown a deviation from strong SOC dominated $J_{eff}$ = 1/2 state in Sr$_2$IrO$_4$.\cite{haskel} Additionally, a recent neutron diffraction measurement has shown moment of Ir in Sr$_2$IrO$_4$ to be 0.208 $\mu_B$ which appears much low compared to the calculated spin-only value 1 $\mu_B$/Ir-site for localized spin of $S$ = 1/2.\cite{ye} While there is no clear experimental evidences for metallic behavior in paramagnetic state of Sr$_2$IrO$_4$, the band structure calculations by Arita \textit{et al.}\cite{arita} have clearly predicted Sr$_2$IrO$_4$ to be a Slater-type insulator. Another calculation by Watanabe \textit{et al.}\cite{watanabe} has shown Sr$_2$IrO$_4$ has moderate electronic correlation effect and basically lies in between Slater- and Mott-type insulating phase. On experimental side, a time-resolved optical study has shown an unique coexistence of Slater- and Mott-type insulating phase in Sr$_2$IrO$_4$.\cite{hsieh} A recent study with scanning tunneling microscopy has interestingly shown that insulating gap in Sr$_2$IrO$_4$ opens up around the magnetic transition and the nature of insulating phase in this material is mostly of Slater-type.\cite{li}

Following this controversy, we have attempted to understand the nature of insulating phase in Sr$_2$IrO$_4$ by means of understanding the nature of magnetic interaction. In that connection, we have estimated the critical exponents, analyzed the thermal demagnetization data and studied the magnetocaloric effect (MCE). The estimated exponents while do not exactly match with the universality classes but values are close to the mean-field interaction model. Analysis of thermal demagnetization data at low temperature, however, imply dual presence of localized and itinerant model of magnetization. The MCE experiment suggest magnetic interaction in Sr$_2$IrO$_4$ is of mean-field type. From these, we infer that insulating phase in Sr$_2$IrO$_4$ has contribution from both Slater and Mott mechanism but it is more close to the Slater one.  
  
\begin{figure}
	\centering
		\includegraphics[width=8cm]{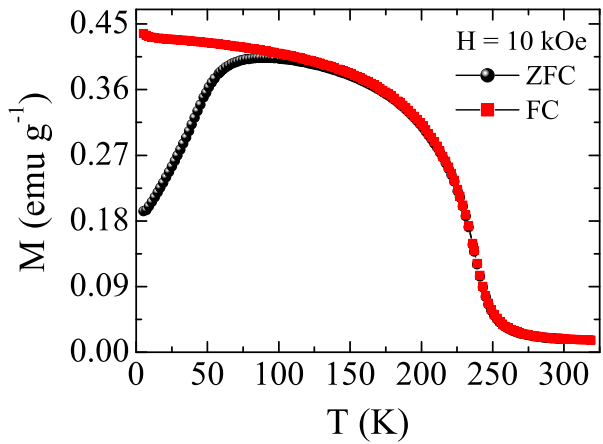}
	\caption{Magnetization measured under ZFC and FC protocol (see text) in applied field of 10 kOe is shown as a function of temperature for Sr$_2$IrO$_4$.}
	\label{fig:Fig1}
\end{figure}

\section{Experimental Methods}
The single-phase polycrystalline sample has been prepared using solid state reaction method. Details of sample preparation and characterization are given in elsewhere.\cite{imtiaz} The magnetization ($M$) data as a function of temperature ($T$) and magnetic field ($H$) have been collected using physical property measurement system (PPMS) by Quantum Design. For critical analysis, magnetic isotherms are collected across $T_c$ in temperature range from 218 to 238 K with a temperature step $\Delta T$ = 2 K. To calculate the change in magnetic entropy $\Delta S_M$, the $M(H)$ data have been collected in temperature range 5 to 300 K up to field 70 kOe.

\section{Results and Discussions}
\subsection{Magnetization Study}
Temperature dependent magnetization data collected following zero field cooled (ZFC) and field cooled (FC) protocol in applied field of 10 kOe are shown in Fig. 1. With decreasing temperature, both branches of magnetization shows steep rise below 250 K which is marked by PM to FM phase transition. At low temperature below about 95 K, the $M_{ZFC}(T)$ shows a sharp fall which has been understood to arise due to complex magneto-structural coupling in this material.\cite{imtiaz} This material otherwise makes transition from PM to AFM state around 230 K. The weak FM behavior is realized a result of spin canting in AFM state which arises due to Dzyaloshinsky-Moriya (DM) type interaction induced by rotation of IrO$_6$ octahedra. In deed, a trace of spontaneous magnetization at low temperature has earlier been observed in Sr$_2$IrO$_4$.\cite{imtiaz}

\begin{figure}
	\centering
		\includegraphics[width=8cm]{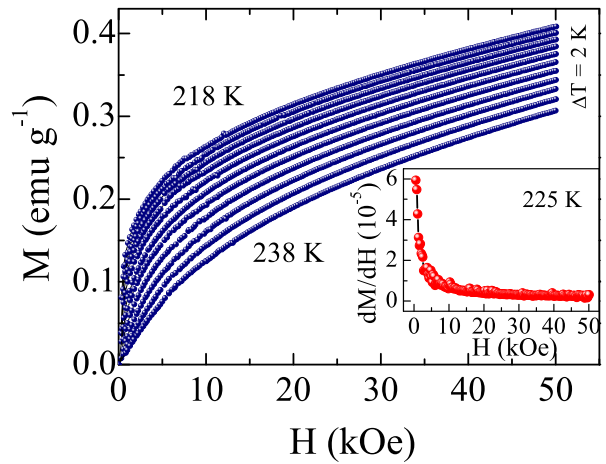}
	\caption{Magnetic isotherms ($M$ vs $H$) measured at different temperatures are shown across $T_c$ for Sr$_2$IrO$_4$.}
	\label{fig:Fig2}
\end{figure}

\subsection{Critical Analysis}
The Sr$_2$IrO$_4$ has magnetic transition from PM to FM state which is of second order. Across the second order phase transition the spontaneous magnetization ($M_S$) below $T_c$, inverse magnetic susceptibility ($\chi^{-1}$) above $T_c$ and the magnetization at $T_c$ follow power-law behavior with temperature as given below:\cite{{stanley}}

\begin{eqnarray}
M_S(T) = M_0(-\epsilon)^\beta, &\epsilon < 0
\end{eqnarray}

\begin{eqnarray}
\chi_0^{-1}(T) = \Gamma(\epsilon)^\gamma, &\epsilon > 0
\end{eqnarray}

\begin{eqnarray}
M = X(H)^{1/\delta}, &\epsilon = 0
\end{eqnarray}

\begin{figure}
	\centering
		\includegraphics[width=8cm]{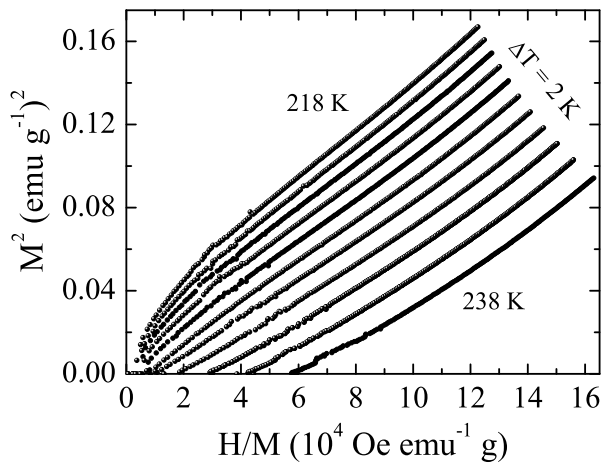}
	\caption{Arrott plot ($M^2$ vs $H/M$) as obtained from isotherms are shown at different temperatures for Sr$_2$IrO$_4$.}
	\label{fig:Fig3}
\end{figure} 

where $\epsilon$ = ($T$ - $T_c$)/$T_c$ is the reduced temperature. The $M_0$, $\Gamma$ and $X$ are the critical amplitudes and $\beta$, $\gamma$ and $\delta$ are the critical exponents. Magnetic phase transitions are generally classified into different universality classes based on the dimensionality of lattice system ($d$) and the dimensionality of spin system ($n$). In principle, critical exponents in asymptotic regime ($\epsilon$ $\rightarrow$ 0) exhibit universal values depending on the universality class the system belong to, where the exponents are independent of microscopic details of the system. To understand the nature of magnetic state in Sr$_2$IrO$_4$ we have estimated the exponents by measuring magnetic isotherms $M(H)$ around the $T_c$. The exponents have been estimated following different methods such as, Arrott plot,\cite{arrott} critical isotherm analysis and Kouvel-Fisher analysis.\cite{kf} Fig. 2 shows $M(H)$ plots measured in temperature range from 218 to 238 K at an interval of 2 K. The general shape of $M$ vs $H$ in main panel and the decreasing slope of $dM/dH$ vs $H$ in inset of Fig. 2 implies a second-order type phase transition in Sr$_2$IrO$_4$. To further analyze the magnetization data we have plotted $M(H)$ isotherms in terms of Arrott plot\cite{arrott} i.e., $M^2$ vs $H/M$. In general, the system which follows mean-field spin interaction model ($\beta$ = 0.5 and $\gamma$ = 1.0) the Arrott plot forms a set of parallel straight lines. The slope taken in higher field regime in Arrott plot directly gives $M_S$ and $\chi_0$ as an intercept on positive $M^2$ and $H/M$ axis, respectively. Moreover, the isotherm which passes through origin in Arrott plot gives $T_c$ of the material.

Fig. 3 shows such Arrott plot of material Sr$_2$IrO$_4$ with the $M(H)$ data from Fig. 2. As clear in Fig. 3, isotherms are not perfectly parallel straight lines and a minor tuning of exponents are required for parallel straight lines. Nonetheless, Fig. 3 suggests the magnetic interaction in Sr$_2$IrO$_4$ is close to mean-field type. It is worth mentioning that we have tried other relevant models for localized spin interaction such as, 3-dimensional Heisenberg and 3-dimensional Ising model but the resulting isotherms do not form set of parallel straight lines.

To find out right values of critical exponents we have used generalized modified Arrott plot:\cite{arrott1}

\begin{eqnarray}
	\left(\frac{H}{M}\right)^{1/\gamma} = a \frac{T - T_c}{T_c} + b M^{1/\beta}
\end{eqnarray}

\begin{figure}
	\centering
		\includegraphics[width=8cm]{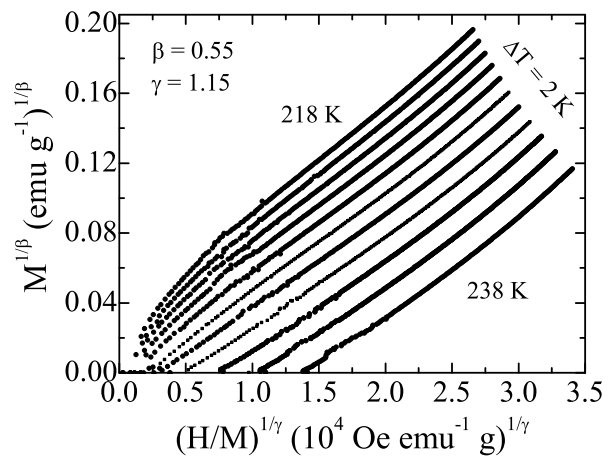}
		\caption{Modified Arrott plot (Eq. 4) constructed from isotherms is shown with exponents $\beta$ = 0.55 and $\gamma$ = 1.15 for Sr$_2$IrO$_4$.}
	\label{fig:Fig4}
\end{figure}

where the $a$ and $b$ are the constants. In Eq. 4, the mean-field exponents $\beta$ = 0.5 and $\gamma$ = 1.0 recover the original Arrott plot. As the Eq. 4 contains two unknown parameters $\beta$ and $\gamma$, therefore tunning these parameters to get parallel straight lines is a nontrivial task. However, we have followed the rigorous iterative process where the $M_S$ and $\chi_0^{-1}$ obtained from Fig. 3 are used in Eqs. 1 and 2 as an input, and obtained new $\beta$ and $\gamma$ values have been used to construct figure similar to Fig. 3. This process has been continued till stable values of $\beta$ and $\gamma$ are obtained. The Fig. 4 shows modified Arrott plot with the so obtained exponents $\beta$ = 0.55 and $\gamma$ = 1.15. It is evident in figure that a set of parallel straight lines is obtained in higher field regime. The lines in lower field regime, however, are curved as the measured magnetization is an average value of contributions from different domains which are magnetized in different directions. We find that isotherm taken at temperature 225 K passes through origin which is the $T_c$ of this material. The obtained $T_c$ ($\sim$ 225 K) in present study conforms with the value determined from neutron diffraction study\cite{ye} (224(2) K) and resonant magnetic x-ray diffuse scattering study\cite{fujiyama} (228.5(0.5) K).  

For further checking, the $M_S$ and $\chi_0^{-1}$ obtained from intercepts on $M^{1/\beta}$ vs $H/M^{1/\gamma}$ axises in Fig. 4, have been plotted in Fig. 5. It is evident in figure that $M_S(T)$ and $\chi_0^{-1}(T)$ reasonably follows Eqs. 1 and 2. The solid lines are due to fitting with Eqs. 1 and 2, respectively. The fitting yields $\beta$ = 0.557(1) and $T_c$ = 225.1 K and $\gamma$ = 1.129(1) and $T_c$ = 224.9 K. The obtained values are pretty close to the values in Fig. 4. 

\begin{figure}
	\centering
		\includegraphics[width=8cm]{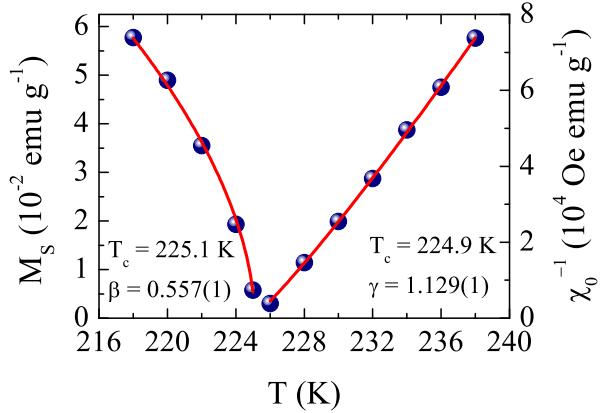}
	\caption{Temperature dependent spontaneous magnetization $M_S$ (left axis) and inverse initial susceptibility $\chi_0^{-1}$ (right axis) determined from the linear fitting of modified Arrott plot in Fig. 4 are shown for Sr$_2$IrO$_4$. Lines are due to fitting with Eqs. 1 and 2.}
	\label{fig:Fig5}
\end{figure}

\begin{figure}
	\centering
		\includegraphics[width=8cm]{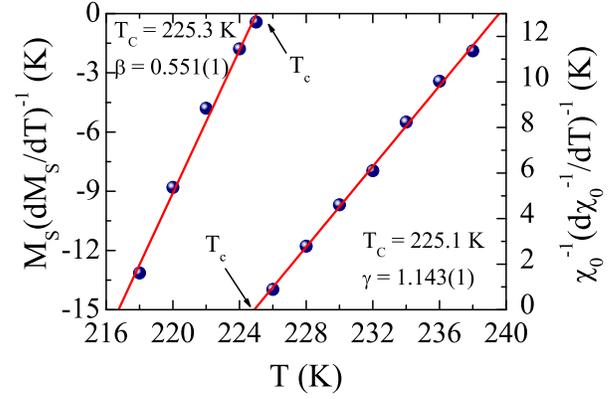}
	\caption{Kouvel-Fisher plots following Eqs. 5 and 6 related to spontaneous magnetization $M_S$ (left axis) and inverse initial susceptibility $\chi_0^{-1}$ (right axis) are shown for Sr$_2$IrO$_4$. Solid lines are due to linear fitting of data.}
	\label{fig:Fig6}
\end{figure}

The critical exponents as well as temperature have also been obtained using Kouvel-Fisher (KF) method.\cite{kf} This method shows following relationships:

\begin{eqnarray}
  M_S\left(\frac{dM_S}{dT}\right)^{-1} = \frac{(T - T_c)}{\beta} 
\end{eqnarray}

\begin{eqnarray}
  \chi_0^{-1}\left(\frac{d\chi_0^{-1}}{dT}\right)^{-1} = \frac{(T - T_c)}{\gamma}
\end{eqnarray}   
\setlength{\tabcolsep}{15pt}
\begin{table*}
\caption{\label{tab:table 1} Table shows the values of critical exponents $\beta$, $\gamma$ and $\delta$ determined in this work following various methods for Sr$_2$IrO$_4$. The theoretical values of exponents for mean-field model are given for comparison.}
\begin{ruledtabular}
\begin{tabular}{cccccc}
Composition &Ref. &Method &$\beta$ &$\gamma$ &$\delta$\\
\hline
Sr$_2$IrO$_4$ &This work &modified Arrott Plot &0.55 &1.15 &3.090$\footnotemark[1]$\\
 &This work &Kouvel-Fisher Method &0.551(1) &1.143(1) &3.074(6)$\footnotemark[1]$\\
 &This work &critical Isotherm & & &3.08(2)\\
Mean-field Theory & & &0.5 &1.0 &3.0\\
\end{tabular}
\end{ruledtabular}
\footnotetext[1]{Calculated following Eq. 7}
\end{table*} 

The Eqs. 5 and 6 imply that slopes obtained from plotting of $M_S(dM_S/dT)^{-1}$ vs $T$ and $\chi_0^{-1}(d\chi_0^{-1}/dT)^{-1}$ vs $T$ will give 1/$\beta$ and 1/$\gamma$, respectively. Moreover, $T_c$ can be correctly and independently obtained from the intersection point on temperature axis. Fig. 6 shows Kouvel-Fisher plot with the $M_S$ and $\chi_0^{-1}$ values obtained from Fig. 4. The Fig. 6 shows a reasonable straight line fitting which yields $\beta$ = 0.551(1) and $T_c$ = 225.3 K and $\gamma$ = 1.143(1) and $T_c$ = 225.1 K. The values obtained following Kouvel-Fisher plot are quite consistent with the ones obtained from modified Arrott plot in Fig. 4.

To determine the exponent $\delta$ we have used Eq. 3 which shows plotting of $M(H)$ at $T_c$ in the form of $\log M$ vs $\log H$ would produce straight line with slope 1/$\delta$. Fig. 7 shows critical isotherm $M(H)$ at $T_c$ = 225 K. Inset shows same plot in log-log scale where a straight line fitting yields $\delta$ = 3.08(2). The obtained $\delta$ is close to the value (3.0) predicted for mean-field interaction model. The exponent $\delta$ has also been obtained using Widom scaling equation which predicts a relationship among the exponents $\beta$, $\gamma$ and $\delta$ as following,\cite{widom}

\begin{eqnarray}
  \delta = 1 + \frac{\gamma}{\beta} 
\end{eqnarray}

Using Eq. 7, the exponent $\delta$ has been calculated with the $\beta$ and $\gamma$ obtained from modified Arrott plot and Kouvel-Fisher plot. Table I shows the calculated values of $\delta$ agree well with the value obtained in Fig. 7, thus confirming the authenticity of the determined exponents.

\begin{figure}
	\centering
		\includegraphics[width=7cm]{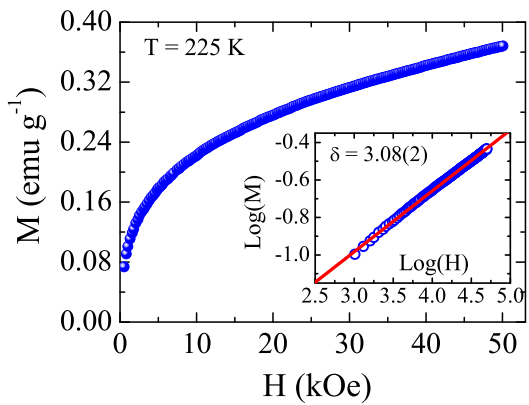}
	\caption{Critical isotherm at 225 K is plotted for Sr$_2$IrO$_4$. Inset shows log-log plotting of same data and the solid line is due to linear fitting.}
	\label{fig:Fig7}
\end{figure}

The critical exponents ($\beta$, $\gamma$ and $\delta$) and the critical temperature ($T_c$) have been obtained from different self-consistent methods and the Table I shows determined values exhibit good agreement. Even though exponents are close to mean-field interaction model, they do not exactly match with the values predicted for established universality classes. We have further checked the consistency of the values using magnetic scaling equation. According to this scaling hypothesis, the magnetization $M(H,\epsilon)$, the magnetic field $H$ and the temperature $T$ obey following relationship,\cite{stanley}

\begin{eqnarray}
	M(H, \epsilon) = \epsilon^{\beta} f_\pm \left( \frac{H}{\epsilon^{\beta + \gamma}}\right)
\end{eqnarray}

where the $f_+ (T > T_c)$ and $f_- (T < T_c)$ are the regular functions. Eq. 8 implies that renormalized magnetization $m$ = $\epsilon^{-\beta} M(H,\epsilon)$ plotted as function of renormalized field $h$ = $\epsilon^{-(\beta + \gamma)}H$ with correct set of exponents and $T_c$ will fall in two distinct branches; one above $T_c$ and another below $T_c$. The main panel of Fig. 8 shows plotting of Eq. 8 and the inset shows the same plot in logarithmic scale. It is evident that renormalized $m(h)$ distinctly fall in two set of branches. Fig. 8 conclusively shows the estimated critical exponents $\beta$, $\gamma$ and $\delta$ and critical temperature $T_c$ are very authentic within the experimental accuracy. To further check the consistency of the determined critical exponents and $T_c$ we have plotted $m$ and $h$ in form of Arrott plot i.e., $m^{2}$ vs $h/m$ as shown in Fig. 9. The figure shows isotherms below and above $T_c$ fall into two branches. This rigorous exercise confirms that the obtained critical exponents and $T_c$ are correct. 

\begin{figure}
	\centering
		\includegraphics[width=8cm]{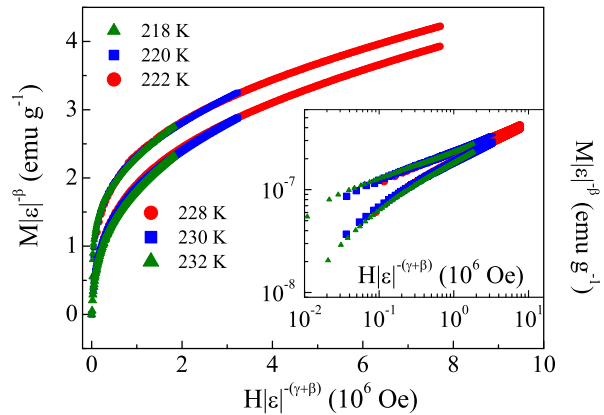}
		\caption{Magnetic isotherms collected at both below and above $T_c$ are plotted following Eq. 8 for Sr$_2$IrO$_4$. The isotherms are clearly scaled into two branches below and above $T_C$ with set of exponents $\beta$ and $\gamma$ and T$_c$.}
	\label{fig:Fig8}
\end{figure}

\subsection{Nature of spin interaction}
The estimated critical exponents of Sr$_2$IrO$_4$ do not exactly follow any theoretical models, however, the values are quite close to mean-field interaction model. Nonetheless, exponent values suggest spin interaction is spatially extended. To further understand the nature of magnetic interaction in this material we have followed a model which suggests that spin interaction decays with spatial distance $r$ as; $J(r)$= $r^{-(d+\sigma)}$ where $d$ is the dimensionality of system and $\sigma$ is the range of interaction.\cite{fisher} According to this model, the $\sigma$ $>$ 2 implies a faster decay of $J(r)$, hence the spin interaction is of short-range type. On the other hand, $\sigma$ $<$ 1.5 is indicative of long-range type interaction. Further, using the renormalization group approach the exponent for magnetic susceptibility $\gamma$ can be calculated for a particular value of $\left\{d:n\right\}$ as following;\cite{pramanik,fischer}

\begin{eqnarray}
 \begin{aligned}
\gamma &= 1 + \frac{d}{4}\frac{(n+2)}{(n+8)} \Delta \sigma + \frac{8(n+2)(n-4)}{d^{2}(n+8)^{2}} \\ 
&\times \left[1 + \frac{2G(\frac{d}{2})(7n+20)}{(n-4)(n+8)}\right] \Delta\sigma^2
\end{aligned}
\end{eqnarray}

where $\Delta\sigma = (\sigma - \frac{d}{2})$ and $G(\frac{d}{2})= 3 - \frac{1}{4} \left(\frac{d}{2}\right)^2$. Following Eq. 9, we have calculated $\gamma$ by varying $\sigma$ for different set of $\left\{d:n\right\}$. For $\sigma$ = 1.13 and $\left\{d:n\right\}$ = $\left\{2:3\right\}$, we obtain $\gamma$ = 1.145 which turns out to be close to the value experimentally determined for present material (see Table I). The $\sigma$ = 1.13 (less than 1.5) implies that the nature of spin interaction in Sr$_2$IrO$_4$ is of long-range type thus supporting the estimated exponent values. The $d$ = 2 indicates 2-dimensional nature of spin interaction which is quite expected considering the layered structure of this material as realized from generalized Ruddlesden-Popper series (SrIrO$_{3}$)$_n$(SrO) with $n$ = 1. In fact, 2-dimensional spin interaction has been concluded from resonant magnetic x-ray diffuse scattering measurements showing an anisotropic magnetic interaction along in-plane and out-plane direction.\cite{fujiyama} The study has shown in-plane spin correlation survives with sizable strength at temperature much higher ($>$ 25 K) than the magnetic ordering temperature 228.5 K and has given an estimate of exchange coupling constant $J$ $\sim$ 0.1 eV. On the other hand, out-plane spin correlation exhibits a critical divergence at ordering temperature and shows much weak coupling constant $J$ $\sim$ $\mu$eV. The $n$ = 3 signifies Heisenberg type spin interaction. Although this is not consistent with the present set of estimated exponents, however, it is likely that short-range interaction coexist with the long-range one which is discussed in net section. 

\begin{figure}
	\centering
		\includegraphics[width=8cm]{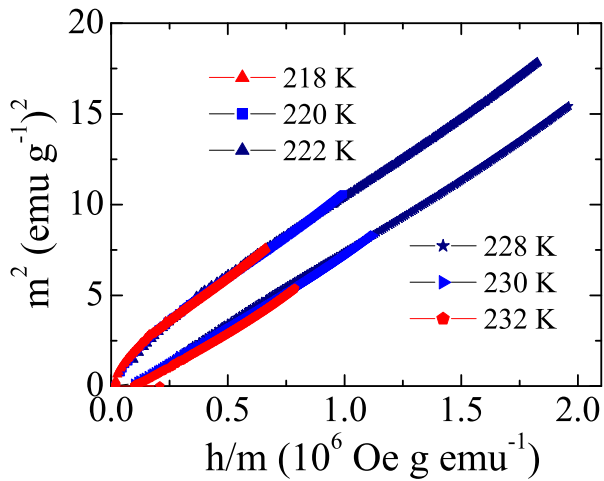}
		\caption{The renormalized $m$ and $h$ are plotted in form of $m^2$ vs $h/m$ for Sr$_2$IrO$_4$.}
	\label{fig:Fig9}
\end{figure}

\subsection{Spin-wave analysis}     
The present material Sr$_2$IrO$_4$ shows weak ferromagnetism arising out of canted type antiferromagnetic ordering which justifies the analysis of critical behavior. In fact, low-frequency ferromagnetic resonance has been evidenced in electron spin resonance (ESR) study for this material \cite{bahr}. With the same spirit, we have attempted to analyze the low temperature thermal demagnetization phenomenon using both spin-wave (SW) and single particle (SP) model. In the picture of localized moment, the thermal demagnetization at low temperature is generally explained with the spin-wave excitations which follows Bloch equation,\cite{das,kaul,kaul1}

\begin{eqnarray}
	\frac{\Delta M}{M(0)} = \frac{M(0) - M(T)}{M(0)} = BT^{3/2} + CT^{5/2} + ........  
\end{eqnarray}

where $B$ and $C$ are the coefficients, $M(0)$ is the magnetization at 0 K. The $T^{3/2}$ term arises due to harmonic contribution and the $T^{5/2}$ term originates from higher order term in spin-wave dispersion relation. The spin-wave stiffness constant $D$ can be calculated as,

\begin{eqnarray}
	D = \frac{k_B}{4\pi} \left[\frac{2.612g\mu_B}{M(0)\rho B}\right]^{2/3}
\end{eqnarray}

where $k_B$ is the Boltzmann constant and $\rho$ is the density of material. In parallel picture of itinerant or band magnetism where the net moment of system is directly proportional to the displacement energy between spin-up and spin-down subbands, the thermal demagnetization is realized as a result of excitation of electrons from one subband to other. The single-particle excitation is generally expressed as,\cite{das}

\begin{eqnarray}
	\frac{\Delta M}{M(0)} = \frac{M(0) - M(T)}{M(0)} = AT^{3/2} \exp\left(-\frac{\Delta}{k_B T}\right)   
\end{eqnarray}
  
where $A$ is the coefficients, $\Delta$ is the energy gap between the top of full sub-band and the Fermi level and k$_B$ is the Boltzmann constant. In a system where the decrease of magnetization with temperature is not significant these processes are quite independent. Thus, the thermal demagnetization can be explained with the contribution from both spin-wave and single-particle excitation.
 
\begin{figure}
	\centering
		\includegraphics[width=8cm]{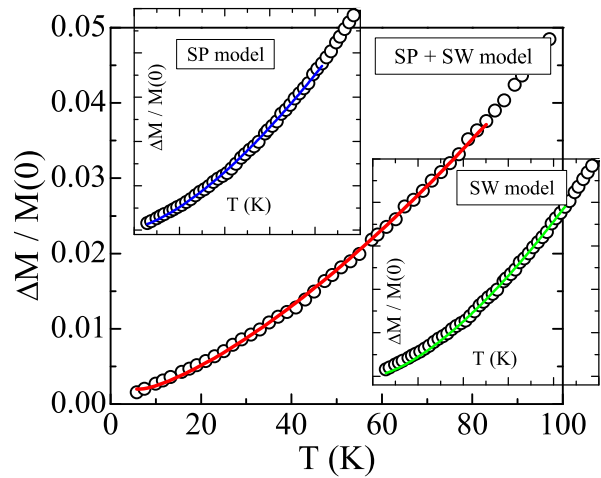}
		\caption{Temperature dependent reduced magnetization is plotted for Sr$_2$IrO$_4$. The lower inset shows fitting of data using spin-wave model (Eq. 10) and upper inset shows the same for single-particle model (Eq. 12). The main panel shows fitting incorporating both model (see text).}
	\label{fig:Fig10}
\end{figure}
       
Fig. 10 shows temperature dependent reduced magnetization $\Delta M$ which is deduced from $M_{FC}$ data (Fig. 1). The $M(0)$ has been estimated from extrapolation of $M(T)$ data. Initially, we tried to fit the data using Eq. 10 with only $T^{3/2}$ dependence, however, it did not yield good fitting (lower inset, Fig 10). The addition of $T^{5/2}$ term also did not improve much the fitting. Similarly, fitting with only SP model using Eq.12 also does not give best result, though it is better than SW model. We found, however, best fitting upon incorporating both SW (up to $T^{3/2}$ term in Eq. 10) and SP model (Eq. 12). The main panel of Fig. 10 shows fitting of data till 0.4$T_c$. The fitting yields $B$ = 3.68 $\times$ 10$^{-5}$ K$^{-3/2}$, $A$ = 1.04 $\times$ 10$^{-5}$ K$^{-5/2}$ and $\Delta$ = 1.159(2) meV. Using Eq. 11, we have calculated stiffness constant $D$ = 3.365(6) eV\AA$^2$. The high value of $D$ can be understood considering canted type antiferromagnetic ordering in this material which gives very low $M(0)$. The low value of gap $\Delta$ = 1.159(2) meV is quite interesting. Note, that similarly small AFM spin excitation gap ($\sim$ 0.85 meV) has been observed in ESR study \cite{bahr}. To understand this a detailed investigation is required. Nonetheless, this analysis primarily shows presence of both localized and itinerant type of magnetization in Sr$_2$IrO$_4$ which is consistent with the fact that estimated exponents do not exactly match with any particular model (Table I). Moreover, simultaneous presence of both Slater and Mott type of insulating phase which is essentially related to itinerant and localized model magnetism, respectively has been previously shown using both experiment and theory for Sr$_2$IrO$_4$ \cite{hsieh,li,watanabe}.

\begin{figure}
	\centering
		\includegraphics[width=8cm]{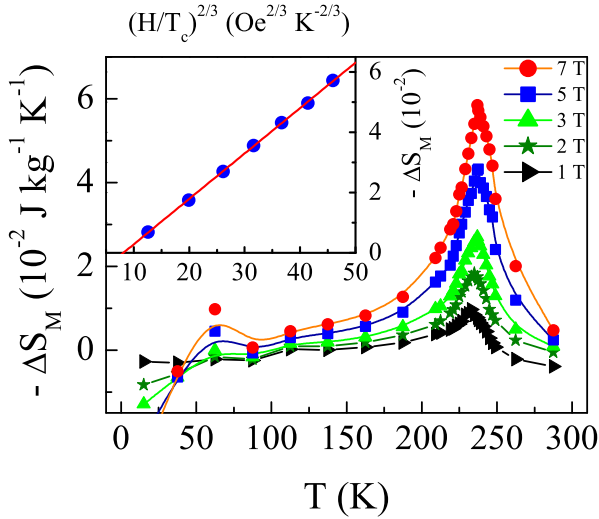}
	\caption{The change in magnetic entropy as calculated using Eq. 15 is shown with temperature for Sr$_2$IrO$_4$. Different plots correspond to different highest applied magnetic fields. Lines are guide to eyes. Inset shows linear dependence of change of magnetic entropy with scaled magnetic field $(H/T_c)^{2/3}$.}
	\label{fig:Fig11}
\end{figure} 

\subsection{Magnetocaloric effect}  
Magnetocaloric effect (MCE) or the change in magnetic entropy ($\Delta S_M$) of a material upon application of magnetic field is a vital tool to understand the magnetic phase transition as well as the magnetic state. The change in entropy with field is related with change of magnetization with temperature through thermodynamic Maxwell relation as,\cite{phana}

\begin{figure}
	\centering
		\includegraphics[width=8cm]{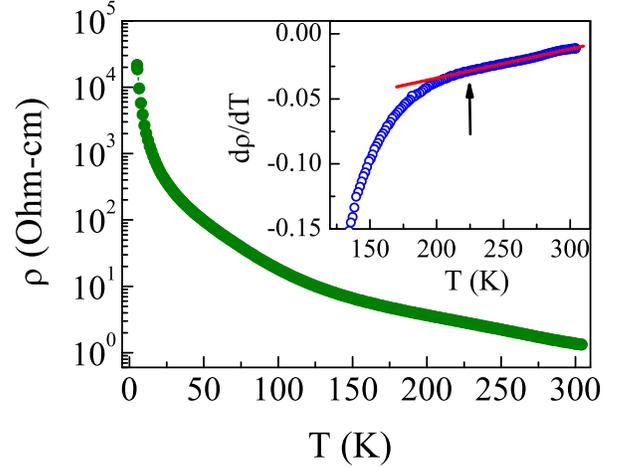}
	\caption{Temperature dependent resistivity is shown for Sr$_2$IrO$_4$. Inset shows derivative of resistivity as a function of temperature.}
	\label{fig:Fig12}
\end{figure}

\begin{eqnarray}
	\left(\frac{\delta S}{\delta H}\right)_T = - \left(\frac{\delta M}{\delta T}\right)_H
\end{eqnarray}
   
The change in magnetic entropy $\Delta S_M(T,H)$ can be calculated as,

\begin{eqnarray}
	\Delta S_M(T,H) = \int_0^H \left(\frac{\delta M(T, H)}{\delta T}\right)_H dH
\end{eqnarray}

When magnetization data is at discrete value of temperature and field intervals, $\Delta S_M(T,H)$ can be calculated using following relation,

\begin{eqnarray}
\begin{aligned}
  \Delta S_M(T,H) = \\ &\sum_i \left(\frac{M_{i+1}(T_{i+1}, H) - M_i(T_{i}, H)}{T_{i+1} -T_i}\right) \Delta H
\end{aligned}
\end{eqnarray}

The $\Delta S_M$ has been calculated for Sr$_2$IrO$_4$ using Eq. 15 from isotherms $M(H)$ collected at different temperatures. The $\Delta S_M$ has been calculated up to different magnetic fields. The $\Delta S_M(T)$ plotted in Fig. 11 shows distinct peak around PM to FM phase transition which is also expected from Eq. 14. At low temperature, where $M_{ZFC}(T)$ shows downfall (see Fig. 1), a weak peak in $\Delta S_M(T)$ is observed in Fig. 11. The calculated $\Delta S_M$ for Sr$_2$IrO$_4$ is rather low which is justified considering the fact that this material is a weak ferromagnetic system. Further, we have used MCE to understand the nature of magnetic state as it is shown that in case of mean-field model, $\Delta S_M$ shows a power law dependence with scaled filed as, $\Delta S_M$ $\propto$ $(H/T_c)^{2/3}$.\cite{oester,dong} The inset of Fig. 11, in deed, shows $\Delta S_M$ varies linearly with $(H/T_c)^{2/3}$ which further confirms magnetic interaction in this material follows mean-field spin interaction model.

\subsection{Discussions}
The estimated critical exponents in Table I imply that the magnetic interaction in Sr$_2$IrO$_4$ do not exactly follow the mean-field model but the nature of interaction is quite close to the mean-field type. The analysis of thermal demagnetization (Fig. 10) suggests that neither localized spin-wave model nor itinerant Stoner single-particle model can fully explain our data. In fact, inclusion of both the model is necessary to understand the thermal demagnetization effect (Fig. 10). Following the similar trend, the MCE or the change in magnetic entropy ($\Delta S_M$) shows a field dependence as, $\left(H/T_c\right)^{2/3}$ which is in conformity with mean-field model (inset of Fig. 11). These experimental observations indicate that the insulating phase in Sr$_2$IrO$_4$ is more close to Slater-type, however, some Mott contribution is also present. This coexistence of both these types of mechanism for insulating phase in Sr$_2$IrO$_4$ has also been captured in other theoretical and experimental investigations.\cite{hsieh,li,watanabe}

While the previous experimental data have some evidences in favor of Slater mechanism the resistivity data, on other hand, have not shown clear metal to insulator transition around magnetic transition till date. In Fig. 12, we have shown resistivity ($\rho$) with temperature for present Sr$_2$IrO$_4$ showing an insulating behavior throughout the temperature range. Interestingly, the temperature derivation of resistivity ($d\rho/dT$) which is plotted in inset shows distinct change in slope across the magnetic transition temperature $T_N$ or $T_c$ which implies a softening of insulating behavior in PM state. In view of Slater model this behavior is not surprising considering the fact that the in-plane exchange interaction in this material is quite robust and survives at temperature much higher than the transition temperature $T_N$,\cite{fujiyama} which probably does not allow the metallic phase in PM state. The simultaneous presence of Mott-type insulating phase may also be responsible for continued insulating phase in PM state.

\section{Conclusion} 
In conclusion, in an aim to understand the nature of insulating phase in Sr$_2$IrO$_4$, whether it is of Slater- or Mott-type, we have estimated critical exponents around the magnetic phase transition, analyzed the thermal demagnetization data and calculated the change in magnetic entropy. The exponent values do not exactly match the values predicted for universality classes, however, the values are quite close to mean-field model. The analysis of exponent using renormalization group approach suggest spin interaction is of 2-dimensional Heisenberg type having extended character. The thermal demagnetization behavior at low temperature can only be explained with inclusion of both localized spin-wave and itinerant Stoner single-particle model. The functional change in magnetic entropy with the applied field is suggestive of mean-field model. We conclude that the insulating phase in Sr$_2$IrO$_4$ is close to Slater-type though some contribution due to Mott-type is also present.
\section{Acknowledgment}
We acknowledge UGC-DAE CSR, Indore for magnetization data. We sincerely thank Alok Banerjee for the magnetization data and discussions. We are also thankful Kranti Kumar for the help in measurements.

\end{document}